\documentclass[pra,aps,showpacs,twocolumn]{revtex4}
\usepackage{amsfonts}
\usepackage{amsmath}
\usepackage{amssymb}
\usepackage[dvips]{graphicx}

\begin{document}

\title{Entanglement properties in the Inhomogeneous Tavis-Cummings model }
\author{C. E. L\'{o}pez}
\author{F. Lastra}
\author{G. Romero}
\author{J. C. Retamal}
\affiliation{Departamento de F\'{\i}sica, Universidad de Santiago de Chile, Casilla 307
Correo 2, Santiago, Chile}
\date{\today}

\begin{abstract}
In this work we study the properties of the atomic entanglement in
the eigenstates spectrum of the inhomogeneous Tavis-Cummings
Model. The inhomogeneity is present in the coupling among the
atoms with quantum electromagnetic field. We calculate analytical
expressions for the concurrence and we found that this exhibits a
strong dependence on the inhomogeneity.
\end{abstract}

\maketitle

\section{Introduction}

Quantum correlations have played a central role from the very
beginning of Quantum Mechanics, and currently the concept of
Entanglement has become a key resource in the research on quantum
information and quantum computation~\cite{Nielsen}. The
availability of entangled quantum states and their
characterization are among the most important questions in quantum
information. In this sense, there have been a considerable number
of theoretical works that have allowed for characterizing quantum
entanglement. From a mathematical point of view, a quantum state
of a pair of particles is separable if this can be written as a
convex sum of product states~\cite {werner}, $\rho
=\sum_{i}^{k}p_{i}|a_{i}\mathbf{\rangle }\left\langle
a_{i}\right\vert \otimes |b_{i}\mathbf{\rangle }\left\langle
b_{i}\right\vert $, where $|a_{i} \mathbf{\rangle }$ and
$|b_{i}\mathbf{ \rangle }$ correspond to different particles in
the pair and $ \sum_{i}p_{i}=1 $. Actually, separability
criteria~\cite{peres,horodecki} provides us with necessary and
sufficient conditions to state if a bipartite quantum state is
separable or not. However, these criteria do not allow us to
quantify exactly the amount of Entanglement of a given system.
Bennet \emph{et.al.}~\cite{bennett} have characterized the
necessary and sufficient channel resources to transmit quantum
states. This leads to a measure of entanglement that is called
\emph{Entanglement of Formation}. In other important work,
Wootters~\cite{Hill,Wootters} has shown that the Entanglement of
formation of an arbitrary state of two qubits is defined by an
exactly calculable quantity called \emph{concurrence}.

On the other hand, recent advances in the manipulation of
collections of atoms have led to the possibility of considering
inhomogeneous coupling, for example, in systems of atomic clouds.
Some contributions have emerged inspired by this fact. For
instance, decoherence of collective atomic states due to the
inhomogeneous coupling between the atoms and external fields has
been studied by Sun \emph{et.al.}~\cite{Sun}. Also, Kuzmich
\emph{et.al.}~\cite{Kuzmich} has pointed out the nonsymmetric
character of the entanglement of multi-atom quantum states due the
inhomogeneity. Recently, decoherence in process of quantum
information storage in atomic clouds has been
studied~\cite{Xiong}, where the phenomenon of electromagnetically
induced transparency (EIT) is present.

Atomic cloud physics could be conveniently described by the Dicke
model~\cite {Dicke}, when considering the interaction of atoms
with light in free space. The Tavis-Cummings (TC)
model~\cite{Tavis}, is suitable when the coupling takes place
inside a cavity. In most applications of the TC model, a constant
coupling between the atoms and the radiation field is assumed.
This simplification is, of course, essential when an analytical
description of the coupled system is
required~\cite{cooperativeeffects,jc1,jc2,review1}, at least when
a small number of atoms is involved. For the case of many atoms,
the dynamics can still profit from the associated $SU(2)$ group
structure and numerical solutions can be found. However, the
situation is drastically different when we consider the more
realistic inhomogeneous coupling. In this case, there is no
possibility to access the Hilbert space in a simple manner,
because all angular momenta representations are mixed along the
dynamics and no analytical approach is known.

The entanglement properties of the ground state for the Dicke
model has recently been studied by authors in
Ref.~\cite{Orszag,Ru}. In such case, when the coupling between
atoms and field is homogeneous, the concurrence between two atoms
is independent of the pair considered due to the symmetry of the
Dicke states. A similar situation should arise in the case of the
Tavis Cummings model. From the point of view of the model
properties, it would be interesting to study the quantum
correlations between atoms under more general conditions, as is
for the case of inhomogeneous coupling. An interesting question to
answer would be, how does the spatial profile of the coupling
reflect on entanglement between atoms?.

In this work we aim to describe the entanglement properties in the
spectrum eigenstates of the TC model~\cite{Tavis} by considering
inhomogeneous coupling of the atoms to the quantum electromagnetic
field. Specifically we study the bipartite atomic concurrence by
tracing out $N-2$ particles. This work is organized as follows: in
Sec. II we present the inhomogeneous Tavis-Cummings (ITC) model
and the properties of the Hilbert space are shown. In Sec. III, we
study the properties of the eigenstates of the model, the
bipartite concurrence between atoms is analyzed for different
number of excitations in the system. In Sec. IV, we present our
concluding remarks.

\section{The Model}

The Hamiltonian describing the inhomogeneous interaction of $N$ atoms with a
single mode of a quantum electromagnetic field in the on resonance regime is
given by $H=H_{0}+H_{I}$ where
\begin{equation}
\begin{tabular}{l}
$H_{0}=a^{\dag }a+S_{z}$, \\
\\
$H_{I}=aJ_{+}+a^{\dag }J_{-}$,
\end{tabular}
\label{ham}
\end{equation}
besides we have taken $\hbar =1$, $S_{z}=\sum_{j=1}^{N}\sigma
_{z}^{j}/2$ and $J_{\pm }=\sum_{j=1}^{N}\kappa _{j}\sigma _{\pm
}^{j}$ are the atomic collective operators with $\kappa
_{j}=g_{j}/w$. The number of excitation operator $a^{\dag}a+S_{z}$
is an integral of motion for the system~\cite{Tavis}. In addition
$\kappa _{j}$ is the inhomogeneous coupling of atoms to the field.
In this work, we will consider $\kappa _{j}=\kappa \sin (\pi
x_{j}/L)$. We notice that due to the inhomogeneous coupling, the
collective atomic operators do not satisfy a closed Lie algebra
$su(2)$, leading to a larger total Hilbert space, so that, the
description in terms of symmetric Dicke states is not allowed
anymore.

\begin{figure}[t]
\includegraphics[width=70mm]{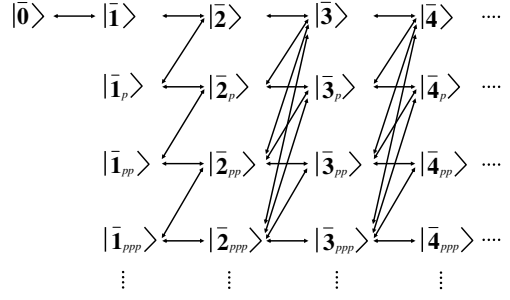}
\caption{Levels scheme that involve the Hilbert space. Here is possible to
see the Hilbert space`s growth due to the Inhomogeneous coupling.}
\label{esquema}
\end{figure}

An effective approach to study the dynamics of systems with
inhomogeneous coupling can be developed~\cite{lrs}. The basic idea
of this approach is to follow the Hilbert space that the coupled
system will visit along the evolution and to implement a
truncation criteria based on a probabilistic argument. In the
homogeneous case the eigenstates of $H$ are built as
superpositions of eigenstates of $H_{0}$, that is, states with a
fixed number of excitations. Thus, the ground state corresponds to
the state with all atoms in the lower state. The first excited
state corresponds to a superposition of one photon state with all
atoms in the ground and the state with zero photons and the
symmetric state with one excitation. In the inhomogeneous case we
have to find which are the states with a fixed number of
excitations. For example, let us consider the state $|n\rangle
|\mathbf{\ \bar{0}\rangle }$ where $|n\rangle $ denotes the state
with $n$ photons and $|\mathbf{\bar{0}\rangle }$ denotes the
collection of $N$ atoms in the ground state $\left\vert
g_{1}g_{2}...g_{N}\right\rangle $. Let us see now how the state
$\left\vert n\right\rangle |\mathbf{\bar{0}\rangle }$ is coupled
to other states under the interaction Hamiltonian~(\ref{ham})
which conserves the number of excitations. We see that the state
$|0\rangle |\mathbf{\bar{0}\rangle }$ is not coupled by the
interaction Hamiltonian to other states of the global system, that
is,
\begin{equation}
|0\rangle |\mathbf{\bar{0}\rangle \longrightarrow }|0\rangle
|\mathbf{\bar{0} \rangle }.
\end{equation}
Now, if we have one excitation in the electromagnetic field
$|1\rangle | \mathbf{\bar{0}\rangle }$ this state is coupled
through the term $J_{+}a$ so that
\begin{equation}
J_{+}a|1\rangle |\mathbf{\bar{0}\rangle =}|0\rangle \sum_{i=1}^{N}\kappa
_{i}|\mathbf{\bar{1}}_{i}\mathbf{\rangle }\text{,}  \label{eq1}
\end{equation}
where $|\mathbf{\bar{1}}_{i}\mathbf{\rangle }$ represents the
state where the $i$-th atom is excited. We define the normalized
state
\begin{equation}
|\mathbf{\bar{1}\rangle =}\frac{1}{\sqrt{\sum_{i=1}^{N}\kappa
_{i}^{2}}}\sum_{i=1}^{N}\kappa
_{i}|\mathbf{\bar{1}}_{i}\mathbf{\rangle }\text{,} \label{one}
\end{equation}
such that the matrix element of the hamiltonian between $|1\rangle
|\mathbf{\bar{0}\rangle }$ and $|0\rangle |\mathbf{\bar{1}\rangle
}$ states is given by
\begin{equation}
\langle 0|\langle \mathbf{\bar{1}}|H|1\rangle
|\mathbf{\bar{0}\rangle =} \sqrt{\sum_{i=1}^{N}\kappa
_{i}^{2}}=N_{1}.
\end{equation}
As we can see, the number of excitations is conserved. The
resulting state is coupled to the state $|1\rangle
|\mathbf{\bar{0}\rangle }$ such that, starting from this root the
system evolves in a closed two dimensional subspace $\{|1\rangle
|\mathbf{\bar{0}\rangle ,}|0\rangle |\mathbf{\bar{1} \rangle }\}$.
If we increase the number of excitations $k$ in the system, for
example $k=2$, one would expect that the Hilbert space will be
spanned by the vectors $\{|2\rangle |\mathbf{\bar{0}\rangle
,}|1\rangle |\mathbf{\ \bar{1}\rangle ,}|0\rangle
|\mathbf{\bar{2}\rangle }\}$, where $|\mathbf{\ \bar{2}\rangle
}\mathbf{=}\frac{1}{N_{2}}2\sum_{i<j}\kappa _{i}\kappa _{j}|
\mathbf{\bar{2}}_{ij}\mathbf{\rangle }$ and
$N_{2}=\sqrt{4\sum_{i<j}\kappa _{i}^{2}\kappa _{j}^{2}}.$ However,
it can be shown that the state $|0\rangle |\mathbf{\bar{2}\rangle
}$ is coupled through $a^{\dag }J_{-}$ to the state $\left\vert
1\right\rangle \left\vert \Phi _{1}\right\rangle ,$ where
$\left\vert \Phi _{1}\right\rangle
=\frac{2}{N_{2}}\sum_{i}(\sum_{j\neq i}\kappa _{i}\kappa
_{j}^{2})|\mathbf{\bar{1}}_{i}\mathbf{\rangle }$, which is
different from $|\mathbf{\bar{1}\rangle }$, so we can say that $
\left\vert \Phi_{1}\right\rangle $ has a component along
$|\mathbf{\bar{1} \rangle }$ and a component along a state
$\left\vert \mathbf{\bar{1}}_{p}\right\rangle $ perpendicular to
the state $|\mathbf{\bar{1}\rangle }$~\cite{lrs}. Thus, if the
number of excitations keep growing, then it is possible to
generate a sequence of collective atomic states and its
respectively perpendicular states as shown by Fig.~\ref{esquema},
where the coupling among atomic collective states are shown.

It is possible to show that the states $\left\vert
\mathbf{\bar{n}}\right\rangle $ (first row in Fig.~\ref{esquema})
can be obtained from the repeated application of the operator
$aJ_{+}$ over the state $\left\vert n\right\rangle \left\vert
\mathbf{\bar{0}}\right\rangle $
\begin{equation}
\left( aJ_{+}\right) ^{n}\left\vert n\right\rangle \left\vert
\mathbf{\bar{0}}\right\rangle =\sqrt{n!}\left\vert 0\right\rangle
\sum_{i_{1}<i_{2}<..<i_{n}}n!\kappa _{i_{1}}\kappa
_{i_{2}}..\kappa
_{i_{n}}|\mathbf{\bar{n}}_{i_{1}i_{2}..i_{n}}\mathbf{\rangle },
\end{equation}
this implies that the collective atomic states $\left\vert
\mathbf{\bar{n}}\right\rangle $ are given by
\begin{equation}
\left\vert \mathbf{\bar{n}}\right\rangle =\frac{1}{N_{n}}
\sum_{i_{1}<i_{2}<..<i_{n}}n!\kappa _{i_{1}}\kappa
_{i_{2}}..\kappa _{i_{n}}|
\mathbf{\bar{n}}_{i_{1}i_{2}..i_{n}}\mathbf{\rangle }\text{,}
\label{nbar}
\end{equation}
with $N_{n}^{2}=\sum\limits_{i_{1}<i_{2}<..<i_{n}}\left( n!\right)
^{2}\kappa _{i_{1}}^{2}\kappa _{i_{2}}^{2}..\kappa _{i_{n}}^{2}$.
On the other hand, the sequence of states $\left\vert
\mathbf{\bar{n}}_{p}\right\rangle $ in the second row of
Fig.~\ref{esquema} are given by
\begin{equation}
\left\vert \mathbf{\bar{n}}_{p}\right\rangle
=\frac{1}{\sqrt{\left\langle \Phi _{n}\mid \Phi _{n}\right\rangle
-\left\vert \left\langle \mathbf{\bar{n}}\mid \Phi
_{n}\right\rangle \right\vert ^{2}}}\left[ \left\vert \Phi
_{n}\right\rangle -\left\langle \mathbf{\bar{n}}\mid \Phi
_{n}\right\rangle \left\vert \mathbf{\bar{n}}\right\rangle \right]
,
\end{equation}
where the states $\left\vert \Phi _{n}\right\rangle $ are written as
\begin{equation}
\left\vert \Phi _{n}\right\rangle
=(n-1)!\sum_{i_{1}<.....<i_{n}}\kappa _{i_{1}}...\kappa
_{i_{n}}\overline{b}_{i_{1}}..._{i_{n}}\left\vert
\mathbf{\bar{n}}_{_{i_{1}}..._{i_{n}}}\right\rangle,
\end{equation}
and
\begin{eqnarray}
\overline{b}_{i_{1}}..._{i_{n}} &=&\sum_{\alpha _{1}\neq i_{1}}\kappa
_{\alpha _{1}}^{2}+...+\sum_{\alpha _{n}\neq i_{n}}\kappa _{\alpha _{n}}^{2}
\notag\\
&=&\left( N-1\right) \sum_{j=1}^{N}\kappa _{j}^{2}.
\end{eqnarray}
An accurate description of the model should consider all the states we have
found associated with the inhomogeneity. However, based on numerical
calculations we have obtained that the spectrum of the Hamiltonian is not
sensitive to perpendicular states further than the first row.
\begin{figure}[t]
\includegraphics[width=70mm]{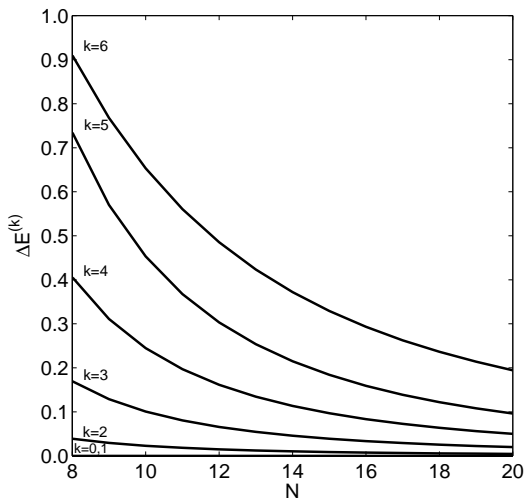}
\caption{Contribution (\%) of second row of Fig.~\ref{esquema} to
the energy as a function of the number of atoms $N$. $\Delta
E^{\left( k\right) }=100(|(E_{1}^{\left( k\right) }-E_{2}^{\left(
k\right) })/E_{1}^{\left( k\right) }|)$, where $E_{1,2}^{\left(
k\right) }$ are the energies for $k$ excitations using $1$ or $2$
rows of Fig.~\ref{esquema}.} \label{deltaE}
\end{figure}
This is shown in Fig.~\ref{deltaE}, where the effects of the
second row on the energies obtained from the first row are
quantified. In this figure we observe that the contribution of the
second row in Fig.~\ref{esquema} is less than $1\%$. Although the
contribution of the second row in Fig.~\ref{esquema} is higher
when the number of excitations increases, decreases when the
number of atoms is increased, as can be seen in Fig.~\ref{deltaE}.
In other words, the contribution of the second row to the energy,
will increase only when $k/N$ increases, where $k$ is the number
of excitations. Thus, we can conclude from Fig.~\ref{deltaE} that
only the first row of Fig.~\ref{esquema} is needed in order to
characterize the spectrum of the ITC model for the parameters used
in this work. On the other hand, it is important to note that this
technique of taking into account only a few rows in
Fig.~\ref{esquema} is also valid when the quantum dynamics of the
system is studied. In that case the approximation is valid within
a given window of time. The size of this window of time will
depend on both $N$ and $k$~\cite{lrs}.

\section{Eigenstates spectrum properties}

In the previous section we have discussed the necessary
ingredients to deal with the ITC model. Particulary, we have shown
that only a part of the Hilbert space associated to the system is
necessary for a suitable description. In this section we study the
properties of the eigenstates of the system, by presenting some
analytical calculations that allow us to characterize the behavior
of the entanglement between a pair of particles. The spectrum and
bipartite atomic Concurrence~\cite{Wootters}, can be analytically
calculated in the inhomogeneous case for $0,1$ and $2$ excitations
respectively. The energies for these cases are given by:
\begin{equation}
\begin{array}{l}
E^{\left( 0\right) }=-\frac{N}{2}\text{,} \\
\multicolumn{1}{c}{} \\
E^{\left( 1\right) }=1-\frac{N}{2}-N_{1}\text{,} \\
\multicolumn{1}{c}{} \\
\multicolumn{1}{c}{E^{\left( 2\right)
}=2-\frac{N}{2}-\sqrt{2N_{1}^{2}+\left( N_{2}/N_{1}\right)
^{2}}\text{,}}
\end{array}
\end{equation}
where $\left( N_{1}\right) ^{2}=\sum_{i=1}^{N}\kappa _{i}^{2}$ and
$\left( N_{2}\right) ^{2}=4\ \sum_{i<j}^{N}\kappa _{i}^{2}\kappa
_{j}^{2}$, correspond to the normalization of the states with one
and two excitations in Eq.~(\ref{nbar}) respectively. These
energies are associated with the eigenstates
\begin{eqnarray}
\left\vert \psi ^{\left( 0\right) }\right\rangle &=&\left\vert
0\right\rangle \left\vert \mathbf{\bar{0}}\right\rangle \text{,}  \notag \\
\left\vert \psi ^{\left( 1\right) }\right\rangle
&=&\frac{1}{\sqrt{2}}\left\{ \left\vert 1\right\rangle \left\vert
\mathbf{\bar{0}}\right\rangle -\left\vert 0\right\rangle
|\mathbf{\bar{1}\rangle }\right\} \text{,}
\label{autoestados} \\
\left\vert \psi ^{\left( 2\right) }\right\rangle &=&\alpha
\left\vert 2\right\rangle \left\vert \mathbf{\bar{0}}\right\rangle
-\frac{1}{\sqrt{2}}\left\vert 1\right\rangle
|\mathbf{\bar{1}\rangle +}\beta \left\vert 0\right\rangle
|\mathbf{\bar{2}\rangle }\text{,}  \notag
\end{eqnarray}
where
\begin{equation}
\begin{tabular}{l}
$\alpha =\sqrt{2}\frac{N_{1}^{2}}{\sqrt{4N_{1}^{4}+2N_{2}^{2}}}$, \\
$\beta =\frac{N_{2}}{\sqrt{4N_{1}^{4}+2N_{2}^{2}}}$.
\end{tabular}
\end{equation}
From the expressions for eigenstates in Eq.~(\ref{autoestados}) we
are able to calculate an analytical expression for the bipartite
concurrence of a pair of atoms $i$, $j$ by tracing out respect to
the other $N-2$ atoms. For this cases the concurrence is given by:
\begin{equation}
\begin{array}{l}
C_{ij}^{\left( 0\right) }=0, \\
\\
C_{ij}^{\left( 1\right) }=\frac{\left\vert \kappa _{i}\kappa _{j}\right\vert
}{\sum_{i=1}^{N}\kappa _{i}^{2}}, \\
\\
C_{ij}^{\left( 2\right) }=2\left\vert \kappa _{i}\kappa
_{j}\right\vert [\frac{1}{2N_{1}^{2}}+\frac{\beta ^{2}}{N_{2}^{2}}M_{1}^{2} \\
\text{ \ \ \ \ \ \ \ \ \ \ }-\frac{\beta }{N_{2}}(\alpha
^{2}+\frac{1}{2N_{1}^{2}}M_{1}^{2}+\frac{\beta
^{2}}{N_{2}^{2}}M_{2}^{2})^{1/2}],
\end{array}
\label{conc1}
\end{equation}
with $M_{1}^{2}=\sum_{k\neq i,j}^{N}\kappa _{k}^{2}$, and $
M_{2}^{2}=\sum_{m<n\neq i,j}^{N}\kappa _{m}^{2}\kappa _{n}^{2}$.

It is important to note from these expressions for the concurrence
that while in the homogeneous TC model the bipartite concurrence
of a pair of atoms is independent of the chosen pair, the
apparition of inhomogeneity in the coupling between the atoms to
the quantum field causes the bipartite concurrence to be dependent
on the pair of atoms that we choose. As is apparent from these
equations, the entanglement between a pair of atoms is
proportional to the coupling constants of atoms to the field. Thus
a pair of atoms located in a region of strong coupling to the
field will be more entangled. As we will see these dependence of
entanglement on the coupling constant appears considering more
excitations.

\begin{figure}[t]
\includegraphics[width=70mm]{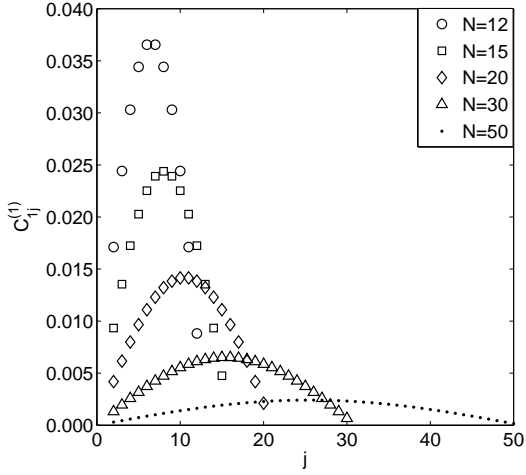}
\caption{Bipartite concurrence $C_{1,j}$ between the first and the $j$-th
atom for $k=1$ excitation for different number of atoms $N$.}
\label{C1j}
\end{figure}

Fig.~\ref{C1j} shows the bipartite concurrence $C_{1,j}$, i.e.,
the concurrence between the first and the $j$-th atom for
different total numbers of atoms $N$. All atoms are equally spaced
for each case. We observe in this picture the effect of the
inhomogeneous coupling of atoms with the field. The concurrence
exhibit a reminiscence of the spatial profile of the single mode
electromagnetic field. The maximum of entanglement happens for the
strongest coupling of the $j$-th atom, which occur for this atom
located at the center the cavity. Comparing the entanglement of
the first atom with the atom located around the center of the
cavity, we realize that as the number of atoms $N$ inside the
cavity is increased, entanglement decreases. This is in agreement
with that found in the homogeneous case.

In the most general case, the state with $k$ excitations can be written as
\begin{equation}
\left\vert \Psi ^{\left( k\right) }\right\rangle
=\sum_{s=0}^{k}A_{s}\left\vert \mathbf{\bar{s}}\right\rangle _{A}\left\vert
k-s\right\rangle _{F}\text{,}
\end{equation}
where $\left\vert \mathbf{\bar{s}}\right\rangle _{A}$ is the
collective atomic state with $s$ excitations defined in
Eq.~(\ref{nbar}) and $A_{s}$ is a coefficient arising from the
eigenvalue problem. Tracing out the quantum field, the reduced
density matrix for the atomic subsystem is given by
\begin{equation}
\rho ^{\left( k\right) }=\sum_{s=0}^{k}\left\vert A_{s}\right\vert
^{2}\left\vert \mathbf{\bar{s}}\right\rangle _{A}\left\langle
\mathbf{\bar{s}}\right\vert.
\end{equation}
Now, tracing out $N-2$ atoms, the bipartite concurrence between atoms $i,j$
\ will be given by
\begin{eqnarray}
C_{i,j}^{\left( k\right) } &=&2\left\vert \kappa _{i}\kappa _{j}\right\vert
[\sum_{s=0}^{k}\frac{\left\vert A_{s}\right\vert ^{2}}{N_{s}^{2}}M_{s-1}^{2}
\notag \\
&&-[\sum_{s,s^{\prime }=0}^{k}\frac{\left\vert A_{s}\right\vert
^{2}\left\vert A_{s^{\prime }}\right\vert ^{2}}{N_{s}^{2}N_{s^{\prime }}^{2}}%
M_{s-2}^{2}M_{s^{\prime }}^{2}]^{1/2}],  \label{Conc}
\end{eqnarray}
where we have defined
\begin{equation}
\begin{tabular}{l}
$N_{p}=\sqrt{\sum\limits_{n_{1}<n_{2}<..<n_{p}}\left( n!\right) ^{2}\kappa
_{n_{1}}^{2}\kappa _{n_{2}}^{2}..\kappa _{n_{p}}^{2}}$, \\
\\
$M_{p}=\sqrt{\sum\limits_{n_{1}<n_{2}<..<n_{p}\neq i,j}\left( n!\right)
^{2}\kappa _{n_{1}}^{2}\kappa _{n_{2}}^{2}..\kappa _{n_{p}}^{2}}$.%
\end{tabular}
\end{equation}

\begin{figure}[t]
\includegraphics[width=70mm]{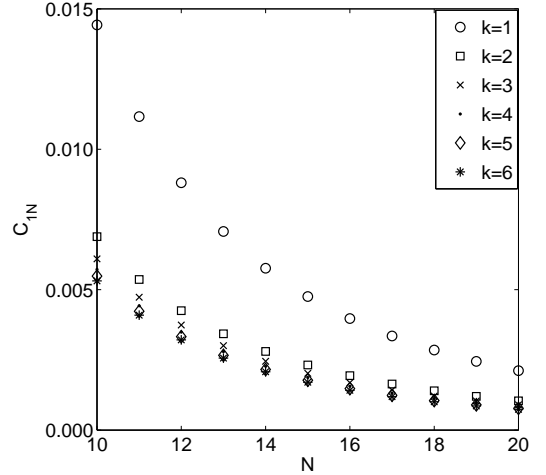}
\caption{Bipartite concurrence $C_{1N}$ between the first and the
last atom for $k=1,2,...,6$ excitations as a function of the
number of atoms $N$.} \label{C1N2}
\end{figure}
\begin{figure}[t]
\includegraphics[width=70mm]{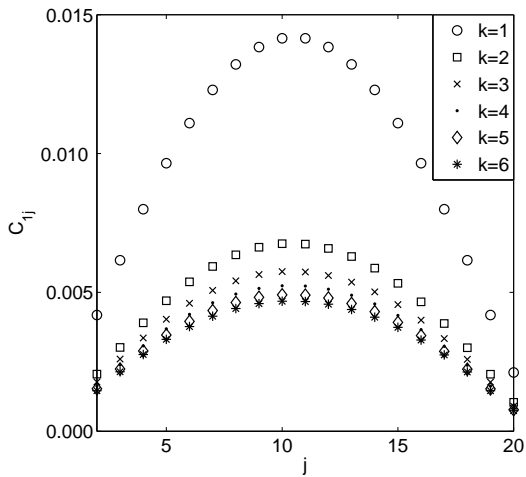}
\caption{Bipartite concurrence $C_{1,j}$ between the first and the
$j$-th atom for $k=1,2,...,6$ excitations with $N=20$ atoms.}
\label{C1jk}
\end{figure}

As we can see in the general expression for the
concurrence~(\ref{Conc}) for $k$ excitations and $N$ atoms, the
entanglement between two of these atoms still depends directly on
the coupling constant. As shown by this result this is one
important feature of entanglement that the TC model exhibit for
the inhomogeneous case. This makes the difference with the the
situation for the homogeneous case where only depends on the total
number of atoms $N$ for a given number of excitations
$k$~\cite{Orszag}. Fig.~\ref{C1N2} shows the concurrence between
the first atom and the last atom as a function of the number of
atoms inside the cavity. In this picture we can observe how the
concurrence decreased when the number of total excitations $k $ in
the system increases. However, when $N\gg 1$ the concurrence
becomes less dependent on the number of excitations tending to a
fixed value. Fig.~\ref{C1jk}, as Fig.~\ref{C1j} shows the
concurrence $C_{1,j}$ between the first and the $j$-th atom in the
cavity. The effects of the inhomogeneous coupling and the loss of
entanglement when the number of excitation is increased are also
clear in this figure.

\section{Summary}

In the present work we have found that although the symmetry
$SU(2)$ no longer exits in the ITC model and consequently a
treatment in terms of symmetric Dicke states is not feasible,
there are still some features that can be analyzed. In particular,
the bipartite concurrence of the eigenstates of the model can be
obtained exhibiting an explicit dependence on the inhomogeneity in
addition to the dependence on the number of atoms and the number
of excitations.

\begin{acknowledgments}
CEL and FL acknowledge the financial support from MECESUP USA0108.
GR from CONICYT Ph. D. Programm Fellowships, and  JCR from
Fondecyt 1030189 and Milenio ICM P02-049.
\end{acknowledgments}

\end{document}